\documentclass[12pt]{iopart}

\input epsf

\begin{document}

\title[Partition function zeros of the one-dimensional Potts model \ldots]
{Partition function zeros of the one-dimensional Potts model: 
The recursive method}

\author{R G Ghulghazaryan\dag\ and\ N S Ananikian\dag\ddag}

\address{\dag\ Department of Theoretical Physics, Yerevan Physics Institute,\\
Alikhanian Brothers 2, 375036 Yerevan, Armenia}
\address{\ddag\
Dipartimento di Scienze Chimiche, Fisiche e Matematiche,\\
Universita Degli Studi Dell'Insubria, via Valleggio,
 11-22100 Como, Italy}
  
\ead{\dag\ \mailto{ghulr@moon.yerphi.am}}
\ead{\ddag\ \mailto{ananik@moon.yerphi.am}}

\begin{abstract}
The Yang-Lee, Fisher and Potts zeros of the one-dimensional Q-state Potts 
model are studied using the theory of dynamical systems. An exact recurrence 
relation for the partition function is derived. It is shown that zeros of the 
partition function may be associated with neutral fixed points of the 
recurrence relation. Further, a general equation for zeros of the partition 
function is found and a classification of the Yang-Lee, Fisher and Potts zeros 
is given. It is shown that the Fisher zeros in a nonzero magnetic field are 
located on several lines in the complex temperature plane and that the number 
of these lines depends on the value of the magnetic field. Analytical expressions 
for the densities of the Yang-Lee, Fisher and Potts 
zeros are derived. It is shown that densities of all types of  zeros of the 
partition function are singular at the edge singularity points with the same 
critical exponent $\sigma=-{\frac12}$. 

\end{abstract}

\submitto{\JPA}
\pacs{0.5.50.+q, 0.5.10.-a,75.10.-b}

\maketitle

\section{Introduction}

It is well established that the thermodynamic properties of a physical system can be derived
from a knowledge of the partition function. Since the discovery of statistical mechanics
it has been a central theme  to understand the mechanism how the 
analytic partition function for a finite-size system acquires a singularity in 
the thermodynamic limit if the system undergoes a phase transition. The answer
to this quest was given in 1952 by Lee and Yang in their famous papers
~\cite{Yang}. They considered the partition function of the Ising model as a 
polynomial in activity ($\exp(-2H/kT)$, where H is a magnetic field) and 
studied the distribution of zeros of the partition function in the complex
activity plane. It was shown that phase transitions occur in the systems where 
continuous distribution of zeros of the partition function cuts the real axis 
in the thermodynamic limit. Also proved, was the circle theorem, which states 
that zeros of the partition function of the ferromagnetic Ising model lie on 
the unit circle 
in the complex activity plane ({\em Yang-Lee zeros}). Later,
Fisher~\cite{Fisher} initiated a study of zeros of the partition function in 
the complex temperature plane ({\em Fisher zeros}).
Fisher showed that complex temperature zeros of the partition function of the
Ising model in zero magnetic field on a square lattice lie on two circles
$|v\pm1|=\sqrt2$, where $v=\tanh(J/2kT)$. Since that time there has been a 
considerable body of work studying the partition function zeros of the Ising 
and Potts~\cite{Wu} models on various regular lattices, most notably in 
recent years ~\cite{other}. Zeros of the partition function were also studied for 
spin models defined on hierarchical~\cite{Derrida} and recursive~\cite{Ghulr} 
lattices, random graphs~\cite{Dolan} and aperiodic systems~\cite{Baake}, 
spin glasses~\cite{Damgaard}, percolation and Self-Organized Criticality 
models~\cite{Arndt}. 

Nowadays, the investigation of zeros of the partition function becomes a 
powerful tool for studying phase transitions and critical phenomena.
Recently, much attention has been attracted to the study of zeros 
of the partition function in one-dimensional systems in connection with 
helix-coil transitions in biological macromolecules~\cite{Alves}. Hence, 
it is very important to study the general properties of zeros of the 
partition functions for different one-dimensional systems.

In 1994, Glumac and Uzelac~\cite{Glumac} using the transfer 
matrix method studied the Yang-Lee zeros of the one-dimensional ferromagnetic 
Potts model for non-integer values of $Q\geq 0$. They showed that for $0<Q<1$ 
the Yang-Lee zeros are located on a real interval and for low temperatures 
these are located partially on the real axis and in complex conjugate 
pairs on the activity plane. Later on, Monroe~\cite{Monroe} 
numerically studied the Yang-Lee zeros of the Potts model for some particular 
values of $Q$. Then, Kim and Creswick~\cite{Creswick} showed that
for $Q>1$ the Yang-Lee zeros lie on a circle with radius $R$, where $R<1$ for
$1<Q<2$, $R>1$ for $Q>2$ and $R=2$ for $Q=2$. Only recently the full 
picture of Yang-Lee zeros of the ferromagnetic Potts model for $0<Q<1$ has been
found~\cite{Ghulr}.

In this paper, the dynamical systems theory is used to study the Yang-Lee, Fisher 
and Potts zeros\footnote{Zeros of the partition function considered as a function 
of complex $Q$.} of the partition function for the one-dimensional $Q$-state Potts 
model. In Section 2, a recurrence relation for the partition function is derived.
It is shown that zeros of the partition function may be associated with neutral
fixed points of the recurrence relation. A general equation for zeros of the 
partition function is derived. Formulae for the free energy and the density of 
zeros of the partition function are found. 
In Sections 3 and 4 the method developed in Section 2 used to study the 
Yang-Lee and Potts zeros of ferromagnetic and antiferromagnetic Potts models. 
A classification of Yang-Lee and Potts zeros is given. 
In Section 5, the Fisher zeros in a nonzero magnetic field are investigated.  
It is shown that the Fisher zeros in a nonzero 
magnetic field located on several lines in the complex temperature plane and
the density of Fisher zeros is singular at the edge singularity points with 
the same critical exponent as that of both the Yang-Lee and Potts zeros, $\sigma=-\frac12$.

\section{Zeros of the partition function of the Potts model}

\begin{figure}[h]
\begin{center}
\epsfxsize=10 cm
\epsfbox{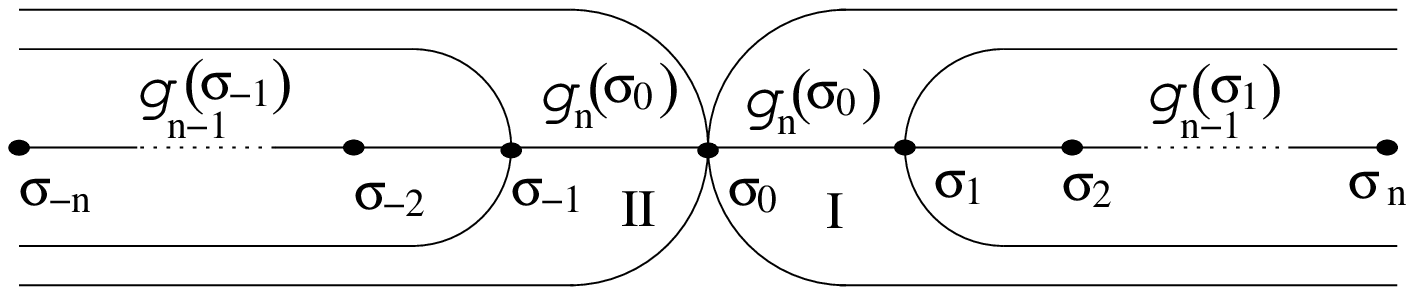}
\end{center}
\caption{\label{1dlattice} The procedure of derivation of the recurrence relation 
for the partition function.}
\end{figure}

The Hamiltonian of the one-dimensional $Q$-state Potts model in a magnetic field 
is defined as follows
\begin{equation}
{\mathcal H}=-\mbox{\~{J}}\sum_{<ij>}
\delta(\sigma_i,\sigma_j)-\mbox{\~{H}}\sum_{i}\delta(\sigma_i,0).
\label{ham}
\end{equation}
where $\delta$ is the Kroneker delta function, $\sigma_i$ denotes the Potts variable 
at site $i$ and takes the values $0,1,2,\ldots,Q-1$. 
The first sum in the r.h.s. of (\ref{ham}) goes over all edges and the
second one over all sites on the lattice. For $\mbox{\~{J}}>0$ the model is 
ferromagnetic and for $\mbox{\~{J}}<0$ is antiferromagnetic. 
Note that due to the symmetry, the Hamiltonian (\ref{ham}) is the same if the 
external field $\mbox{\~{H}}$ is applied to any spin state, namely, 
if $\delta(\sigma_i,0)$ in 
(\ref{ham}) is replaced by $\delta(\sigma_i,\alpha)$ for any $\alpha=1,\,2,\ldots,Q-1$.
For $Q=2$ the Potts model corresponds to the Ising model and in order to keep the analogy
with the Ising model we designate $\mbox{\~{H}}$ as a magnetic field.
We may assume the cyclic boundary condition $\sigma_n=\sigma_{-n}$ and that 
the number of sites is $2n+1$ without losing a generality. Cutting the lattice
at the central site $\sigma_0$ will separate it into two branches I and II with 
equal statistical weights  $g_n(\sigma_0)$ (Figure \ref{1dlattice}).

Cutting the branch I (II) at the site $\sigma_1$ ($\sigma_{-1}$), the 
recurrence relation for $g_n(\sigma)$ may be found as
\begin{equation}
g_n(\sigma_0)=\sum_{\sigma_1=0}^{Q-1}\exp\left[J\delta(\sigma_0,\sigma_1)+%
h\delta(\sigma_1,0)\right]\,g_{n-1}(\sigma_1).
\label{grec}
\end{equation}
where $J=\mbox{\~{J}}/kT$ and $h=\mbox{\~{H}}/kT$. The partition function may be written 
in the form
\begin{equation}
{\mathcal Z}=\sum_{{\sigma}}\rme^{-{\mathcal H}/kT}=\sum_{\sigma_0=0}^{Q-1}%
\exp\left[h\delta(\sigma_0,0)\right]g_n^2(\sigma_0).
\end{equation}
Introducing the notation 
\begin{equation}
x_n=\frac{g_n(\sigma \neq 0)}{g_n(\sigma =0)},  \label{xn}
\end{equation}
and using~(\ref{grec}), the recurrence relation for $x_n$ may be found as
\begin{equation}
x_n=f(x_{n-1}),\qquad f(x)=\frac{\mu +(z+Q-2)\,x}%
{z\,\mu+(Q-1)\,x}, 
\label{xrec}
\end{equation}
where $\mu =\rme^h$, and $z=\rme^J$.
$x_n$ does not have a direct physical meaning, but the 
thermodynamic functions, such as the magnetization, the specific heat and etc., 
may be expressed in terms of $x_n$. For example, the magnetization for our
model in the thermodynamic limit has the form
\begin{equation}
m=(N\,{\mathcal Z})^{-1}\sum_{\{\sigma\}}\delta(\sigma,0)\,\rme^{-{\mathcal H}/kT}%
=\frac{\mu}{\mu+(Q-1)\,x},
\label{magn}
\end{equation}
where $x$ is an attracting fixed point of the mapping (\ref{xrec}). 
It will be shown below how the thermodynamic properties of the model may be
defined from the dynamics of the recurrence relation (\ref{xrec}).

The mapping (\ref{xrec}) is a M\"{o}bius transformation, i.e. a rational map
of the form
\[
R(x)=\frac{a\,x+b}{c\,x+d},\qquad ad-bc\neq0,
\]
where
\[
R(\infty)=a/c, \qquad R(-d/c)=\infty,
\]
if $c\neq 0$, while $R(\infty)=\infty$ when $c=0$. The dynamics of such
maps is well studied~\cite{Beardon}. The mapping (\ref{xrec}) has only 
two fixed points which are solutions to the equation $f(x)=x$. According to
the theory of complex dynamical systems, fixed points are classified
as follows: A fixed point $x^*$ is attracting if $|\lambda|<1$, repelling
if $|\lambda|>1$, and neutral if $|\lambda|=1$, where 
$\lambda={\rmd\over\rmd x}f(x)|_{x=x^*}\equiv f^\prime(x^*)$ is called the 
eigenvalue of $x^*$. 
It is easy to show that either  both fixed points of the mapping (\ref{xrec})
are neutral, or that one of them is attracting while the other is repelling. 
The correspondence between thermodynamical properties of the model and the 
dynamics of the mapping (\ref{xrec}) is the following: If for a given 
temperature and magnetic field ($z$ and $\mu$) the mapping (\ref{xrec}) has an attracting 
fixed point, then the system is in a stable state and its thermodynamical
functions are defined by this fixed point (see for example (\ref{magn})). 
The other fixed point is repelling and does not correspond to any phase. 
On the other hand, if the iterations of (\ref{xrec}) 
do not converge to a fixed point, i.e. the mapping (\ref{xrec}) has neutral 
fixed points only, the system undergoes a phase transition. Therefore, the 
existence of neutral fixed points of (\ref{xrec}) corresponds to a phase 
transition in the model. It was mentioned in the Introduction that zeros 
of the partition function correspond to phase transitions in the model. Thus,
zeros of the partition function are associated with neutral fixed points of 
the corresponding mapping. These may be found from the conditions of existence 
of neutral fixed points of the mapping (\ref{xrec}). 
These conditions are the following
\begin{equation}
\left\{
\begin{array}{l}
f(x)=x \\ 
f^\prime(x)=e^{i\phi},\quad \phi\in [0,2\pi],
\label{1dneutral}
\end{array}
\right.
\end{equation}
Excluding $x$ from the system (\ref{1dneutral}) after
some algebra the equation of phase transitions may be found 
\begin{equation}
z^2\,\mu^2-2\,\left[(z-1)\,(z+Q-1)\,\cos\phi+1-Q\right]\,\mu+\left(z+Q-2\right)^2=0,
\label{phasetr}
\end{equation}
where $\phi\in [0,2\pi]$. Solutions to this equation for different values of $\phi$
correspond to zeros of the partition function. Hence, the free energy of the model 
may be written in the form
\begin{equation}
\fl
F\sim \int_0^{2\pi} \ln(z^2\,\mu^2+2\,(Q-1)\,\mu+\left(z+Q-2\right)^2-%
2\,\mu\,(z-1)\,(z+Q-1)\,\cos\phi\,)\,\rmd \phi.
\label{fenergy}
\end{equation} 
Formula (\ref{fenergy}) may also be derived from the transfer matrix method
~\cite{Glumac,Wu2} using two non-degenerate eigenvalues of the transfer matrix
~\cite{Wu2}
\begin{equation}
\lambda^*_{1,2}=\frac12\left[ z\,\mu+z+Q-2\pm\sqrt{(z\,\mu-z-Q+2)^2+%
4\,(Q-1)\,\mu}\right],
\label{eigen}
\end{equation}
and the following mathematical identity for any pair of scalars $C$, $D$
\[
C^N+D^N=\prod_n\left[C+\exp(2\pi\rmi n/N)\,D\right],
\]
where the product is from $n=1,\,2,\ldots,N$ if $N$ is odd; and from $n=1/2,%
\,3/2,\ldots,N-1/2$ for N even. Moreover, one can easily show that the eigenvalues 
of fixed points of (\ref{xrec}) are related to the eigenvalues of transfer matrix%
~(\ref{eigen}) as follows
\[
\lambda_{1,2}=\frac{4\,\mu\,(z-1)(z+Q-1)}{(\lambda^{*}_{1,2})^2}.
\]
Hence, the condition of a phase transition  $|\lambda_1|=|\lambda_2|$ (both 
$\lambda_{1}$ and  $\lambda_{2}$ are neutral fixed points) corresponds 
exactly to the $|\lambda^*_1|=|\lambda^*_2|$ condition for the eigenvalues
of transfer matrix~(\ref{eigen}). It means that the phase transition point based on 
dynamical systems approach coincides exactly with the phase transition point based 
on free energy 
considerations. This correspondence seems to be general for spin models defined on 
recursive lattices~\cite{Ghulr,Monroe2}.

The density of zeros of the partition function may be found from the equation 
(\ref{phasetr}) in a standard way and has the form
\begin{equation}
g(\xi)=\frac{B\,\partial_\xi A - A\,\partial_\xi B}{2\,\pi\,B\,[-(A-B)(A+B)]^{\frac12}}\,,
\label{density}
\end{equation}
where $\partial_\xi=\frac{\partial}{\partial \xi}$,
\[
\fl
A=z^2\,\mu^2+2\,(Q-1)\,\mu+\left(z+Q-2\right)^2,
\quad
B=2\,\mu\,(z-1)\,(z+Q-1)\,\cos\phi,
\]
and $\xi=\mu,\,z$ or $Q$ depending on whether the
Yang-Lee, Fisher or Potts zeros are considered. It is interesting to note that 
in our case the numerator of (\ref{density}) always contains the multiplier $(A+B)$. 
Hence, the density $g(\xi)$ is singular only when $A-B=0$\footnote{Solutions to
the equation B=0 should be neglected since these do not correspond to zeros of the 
partition function.}. From (\ref{density}) follows that $g(\xi)$ has a singular behavior 
$g(\xi)\sim|\xi-\xi^*|^\sigma$, where $\xi^*$ is a solution to the equation $A-B=0$.
We will see that $\sigma={-\frac12}$ for all types of zeros of the
partition function. Also, note that the equation $A-B=0$ corresponds
 to the phase transitions equation (\ref{phasetr}) for $\phi=0$. It
defines the edge singularity points~\cite{Ghulr,Kortman}. In the subsequent Sections we 
will apply equations (\ref{phasetr}) and (\ref{density}) to the study of the Yang-Lee, 
Potts and Fisher zeros.   

\section{The Yang-Lee zeros}
According to the results of the previous section the Yang-Lee zeros of the
$Q$-state Potts model may be found by solving the equation (\ref{phasetr}) with 
respect to $\mu$. Equation (\ref{phasetr}) is a quadratic equation of $\mu$ with 
real coefficients. Note that solutions to the equation (\ref{phasetr}) lie either 
on the real axis or in complex conjugate pairs on a circle with radius $R=|z+Q-2|/z$ 
and have the form
\begin{equation}
\mu_{1,2}=E\,\Big[2\,\cos^2{\textstyle\frac{\phi}{2}}-F\pm2\,\sqrt{\cos^2%
{\textstyle\frac{\phi}{2}}%
\left(\cos^2{\textstyle\frac{\phi}{2}}-F\right)}\Big], \label{sol}
\end{equation}
where
\[
E=\frac{(z-1)\,(z+Q-1)}{z^2} \quad\mbox{and}\quad
F=\frac{z(z+Q-2)}{(z-1)\,(z+Q-1)}.
\]
A detailed study of (\ref{phasetr}) with respect to $\mu$ has already been performed
~\cite{Ghulr}. Here, we will present only the main results for the Yang-Lee 
zeros of both ferromagnetic and antiferromagnetic Potts models. 

For the 
ferromagnetic Potts model ($z>1$) one can find that for $Q>1$ all solutions 
(\ref{sol}) are complex conjugate and lie on an arc of circle with radius
$R=(z+Q-2)/z$.  Writing $\mu$ in the exponential form $\mu=R\,e^{i\,\theta}$ 
the angular distribution of the Yang-Lee zeros may be found in the form
\begin{equation}
\cos\frac{\theta}{2}=\sqrt\frac{(z-1)\,(z+Q-1)}{z\,(z+Q-2)}\,\cos\frac{\phi}{2}.
\label{ylgapf}
\end{equation}
From (\ref{ylgapf}) one can see a gap in the distribution of Yang-Lee zeros, i.e.
there are no solutions to the equation (\ref{phasetr}) in the interval 
$-\theta_0<\theta<\theta_0$, where $\theta_0=2\,\arccos \sqrt{F^{-1}}$. This is 
the well known gap in the distribution of Yang-Lee zeros of ferromagnetic models
above the critical temperature (formally $T_c=0$ for the one-dimensional case)
first studied by Kortman, Griffiths and Fisher~\cite{Kortman}. 
The end points of the gap are called Yang-Lee edge singularity points.
From (\ref{phasetr}) and (\ref{ylgapf}) follows that the Yang-Lee edge singularity 
points correspond to $\phi=0$. 
Hence, according to (\ref{density}) the density of Yang-Lee 
zeros is singular in the Yang-Lee edge singularity points. Substituting $\phi=0$ into 
(\ref{sol}) the formula for Yang-Lee edge singularity points we have
\begin{equation}
\mu_\pm=\frac{1}{z^2}\left\{\sqrt{(z-1)(z+Q-1)}\pm\sqrt{1-Q}\right\}^2.
\label{edge}
\end{equation}
Substituting $\xi=\mu\equiv R\rme^{\rmi\theta}$ in (\ref{density}) after 
some algebra, the density of Yang-Lee zeros for $Q>1$ may be found in the form
\begin{equation}
g(\theta)=\frac{1}{2\,\pi}\frac{|\sin\frac{\theta}{2}|}{\sqrt{\sin^2\frac{\theta}{2}
 - \sin^2\frac{\theta_0}{2}}},
\label{ylden1f}
\end{equation}
where $\theta_0=2\,\arccos \sqrt{F^{-1}}$.
From the equation (\ref{ylden1f}) follows that the density $g(\theta)$ 
diverges in the Yang-Lee edge singularity points $\mu_\pm$ with the critical exponent 
$\sigma=-\frac 12$, i.e. $g(\theta)\propto |\theta-\theta_0|^{-\frac12}$ when 
$\phi\to 0$ or $\theta\to\theta_0$.

\begin{figure}[h]
\epsfxsize=14cm
\epsfbox{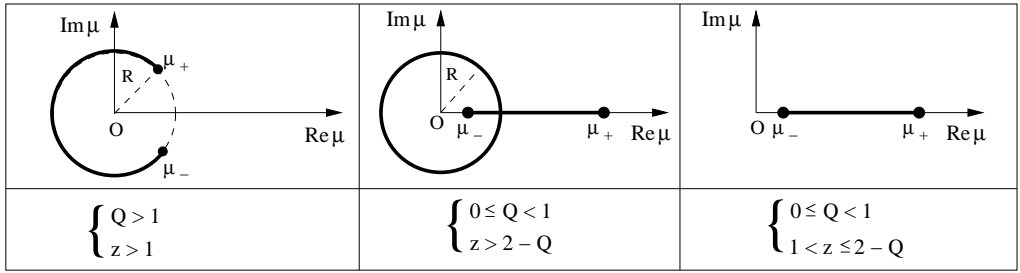}
\caption{A schematic presentation of the Yang-Lee zeros of 1D ferromagnetic
Potts model. Here $R=\frac{z+Q-2}{z}$ and $\mu_\pm$ are defined in (\ref{edge}).
\label{1d:ferro}}
\end{figure}

For $Q<1$ the Yang-Lee edge singularity points $\mu_\pm$ are real and the density 
of Yang-Lee zeros has the form
\begin{equation}
g(\mu)=\frac{1}{2\,\pi\,\mu}\frac{|\mu-\sqrt{\mu_+\mu_-}|}%
{\sqrt{(\mu_+-\mu)(\mu-\mu_-)}}.
\label{ylden2f}
\end{equation}  
$g(\mu)$ diverges in the points $\mu_\pm$, i.e. 
$g(\mu)\propto |\mu-\mu_\pm|^{\sigma}$, with the critical exponent 
$\sigma=-\frac12$. The summary of results  
for the Yang-Lee zeros of the ferromagnetic Potts model is given in Figure
~\ref{1d:ferro}. 

\begin{figure}[h]
\epsfxsize=14cm
\epsfbox{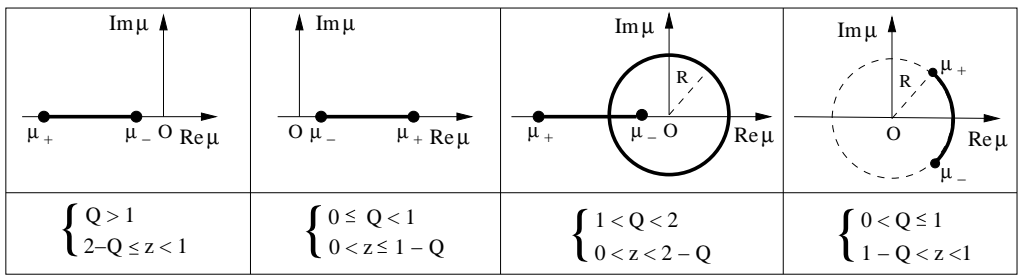}
\caption{A schematic presentation of the Yang-Lee zeros of 1D
antiferromagnetic Potts model. Here $R=\frac{2-Q-z}{z}$ and $\mu_\pm$ are 
defined in (\ref{edge}). \label{1d:antiferro}}
\end{figure}

The Yang-Lee zeros of the antiferromagnetic Potts model ($z<1$) 
may be studied in the same manner and the results are given in Figure
~\ref{1d:antiferro}. For complex $\mu_\pm$ the angular distribution of Yang-Lee
zeros has the form 
\begin{equation}
\sin\frac{\theta}{2}=\sqrt\frac{(z-1)\,(z+Q-1)}{z\,(z+Q-2)}\,\cos\frac{\phi}{2}.
\label{ylgapaf}
\end{equation}
In contrast to the ferromagnetic case now the Yang-Lee zeros lie in the interval 
$-\theta_0<\theta<\theta_0$, where $\theta_0=2\arcsin\sqrt{F^{-1}}$.
The density of Yang-Lee zeros in this case has the form
\begin{equation}
g(\theta)=\frac{1}{2\,\pi}\frac{\cos\frac{\theta}{2}}%
{\sqrt{\sin^2\frac{\theta_0}{2} - \sin^2\frac{\theta}{2}}}.
\label{ylden1af}
\end{equation}
For real values of $\mu_\pm$ the density of Potts zeros has the same form as for 
the ferromagnetic case~(\ref{ylden2f}). 
Actually, the formula (\ref{ylden2f}) is the most general from which formulae
(\ref{ylden1f}) and (\ref{ylden1af}) may be derived (See also Section 4). 
Note, that here also the density of Yang-Lee zeros diverges at the Yang-Lee edge
singularity points with the same index $\sigma=-\frac12$.

\section{The Potts zeros}

Recently, much attention has been attracted to the study of zeros of the 
partition function in the complex $Q$ plane for the Potts model on regular 
lattices and finite graphs~\cite{Sokal}. 
The partition function of the 
Potts model ($H=0$) on a finite graph $G$, ${\mathcal Z}_G(Q,v)$, may be considered
as a polynomial in the number $Q$ of Potts states 
and the temperature-like variable $v=\exp(\mbox{\~{J}}/kT)-1$. 
At zero-temperature the partition function of the antiferromagnetic Potts model 
($v=-1$) correspond to the chromatic polynomial $P_G(Q)$, which is closely related
to the $Q$-coloring problem. By definition,  for graph $G$ and positive $Q$, $P_G(Q)$
is  the number of ways in which the vertices of $G$ can be assigned ``colors'' from the 
set $1,2,\ldots, Q$ in such a way that adjacent vertices always receive different
colors. The original hope was that study of the real or complex zeros of $P_G(Q)$
might lead to an analytic proof of the Four-Color Conjecture, which states that 
$P_G(4)>0$ for all loop-less planar graphs. To date this hope has not been realized, 
although combinatoric proofs of the Four-Color Theorem have been found~\cite{Appel}.
Even so, the Potts zeros and zeros of $P_G(Q)$  are interesting in their own right 
and have been extensively studied in recent years~\cite{Sokal, Shrock}.

\begin{figure}[h]
\epsfxsize=14cm
\epsfbox{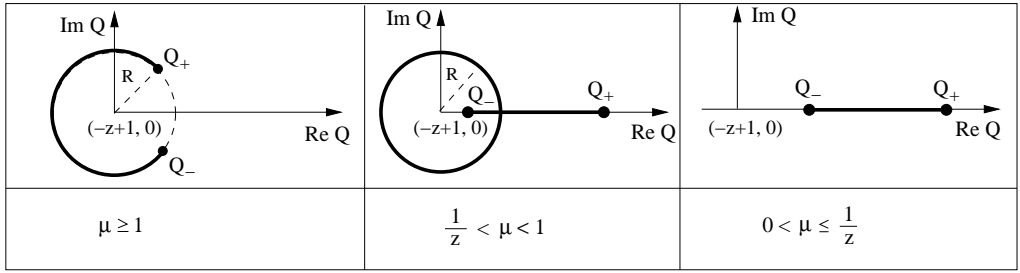}
\caption{A schematic presentation of the Potts zeros of 1D ferromagnetic
Potts model. Here $R=z\,\mu-1$ and $Q_\pm$ are defined in (\ref{Qedge}).
\label{1d:pferro}}
\end{figure}

\begin{figure}[h]
\epsfxsize=14cm
\epsfbox{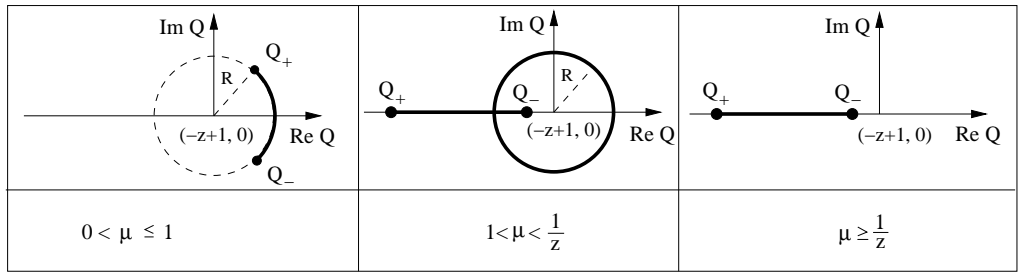}
\caption{A schematic presentation of the Potts zeros of 1D
antiferromagnetic Potts model. Here $R=1-z\,\mu$ and $Q_\pm$ are 
defined in (\ref{Qedge}). \label{1d:pantiferro}}
\end{figure}

In this section the Potts zeros of ferromagnetic and antiferromagnetic
one-dimensional Potts models are studied. 
It was shown in Section 2 that zeros of the partition function correspond to
solutions of the equation (\ref{phasetr}), which is a polynomial in $Q$. 
Hence, the Potts zeros may be found as solutions to (\ref{phasetr})
with respect to the parameter $Q$. It is convenient to rewrite (\ref{phasetr}) using the 
variable $P=Q+z-1$
\begin{equation}
P^2-2\,P\,\left[1-\mu+\mu\,(z-1)\,\cos\phi\right]+(1-z\,\mu)^2=0.
\label{pzeros}
\end{equation} 
Equation (\ref{pzeros}) is a quadratic equation in $P$ with real coefficients.
Solutions of (\ref{pzeros}) lie either on the real axis or on the circle with 
radius $R=|1-z\,\mu|$ in the complex plane $P$, and have the form
\begin{equation}
\fl
P_{1,2}=1-\mu+\mu\,(z-1)\,\cos\phi\pm2\,\cos{\textstyle\frac{\phi}{2}}
\,\sqrt{\mu\,(z-1)(1-\mu-\mu\,(z-1)\sin^2{\textstyle\frac{\phi}{2}})}.
\label{psol}
\end{equation} 
Since the analysis of (\ref{psol}) is standard, we will skip the details and give 
the results in Figures \ref{1d:pferro} and \ref{1d:pantiferro} for ferromagnetic and 
antiferromagnetic Potts models respectively. 
In analogy with the Yang-Lee zeros, the Potts edge singularity points are defined
as solutions to the equation (\ref{pzeros}) for $\phi=0$, and have the form
\begin{equation}
Q_\pm=P_\pm-z+1=1-z+\left(\sqrt{1-\mu}\pm\sqrt{\mu\,(z-1)}\right)^2
\label{Qedge}
\end{equation}

The general formula of the density of Potts zeros has the from
\begin{equation}
g(P)=\frac{1}{2\,\pi\,P}\frac{|P-\sqrt{P_+P_-}|}%
{\sqrt{(P_+-P)(P-P_-)}}.
\label{pdensity2}
\end{equation}  
From (\ref{pdensity2}) it follows that the density of Potts zeros is singular at the 
Potts edge singularity points.
For complex values of $Q_\pm$, (\ref{pdensity2}) may be written in the form
\begin{equation}
g(\theta)=\frac{1}{2\,\pi}\frac{|\sin\frac{\theta}{2}|}%
{\sqrt{\sin^2\frac{\theta}{2} - \sin^2\frac{\theta_0}{2}}},
\label{pden1f}
\end{equation}
for the ferromagnetic model, where $|\theta|>\theta_0$ and 
$\theta_0=2\,\arccos(\mu\,(z-1)/(\mu\,z-1))^{-\frac12}$.
For the antiferromagentic model, $g(\theta)$ has the form
\begin{equation}
g(\theta)=\frac{1}{2\,\pi}\frac{\cos\frac{\theta}{2}}%
{\sqrt{\sin^2\frac{\theta_0}{2} - \sin^2\frac{\theta}{2}}},
\label{pden1af}
\end{equation}
where $|\theta|<\theta_0$ and 
$\theta_0=2\,\arcsin(\mu\,(1-z)/(1-\mu\,z))^{-\frac12}$. 
Different formulae for the density of Potts zeros occur because the
square root function in (\ref{pdensity2}) is not a unique function in 
the complex plane $P$, i.e. one branch of it corresponds to the ferromagnetic 
model and the other to that of the antiferromagnetic. 

It is noteworthy that the density of Potts zeros diverges at the Potts 
edge singularity points with the same critical exponent $\sigma=-\frac12$
as for the Yang-Lee zeros. 

\section{The Fisher zeros}
In this section the Fisher zeros in a nonzero magnetic field are studied.
Usually, the Fisher zeros are considered as zeros of the partition function 
with respect to a temperature-dependent parameter. In our case $z$ is such 
a parameter. Formally, the magnetic field in
the Hamiltonian (\ref{ham}) may be presented in the form $\mbox{\~{H}}=%
H\,\mbox{\~{J}}$, where $H$ is the renormalized magnetic field. Later we
will refer to $H$ as a magnetic field. Then, $\mu$ 
may be written as $z^H$ and the equation (\ref{phasetr}) will has the form  
\begin{equation}
P_\phi(z,H,Q)=0,
\label{fzeros}
\end{equation}
where
\[
\fl
P_\phi(z,H,Q)=z^{2H+2}-2\cos\phi\,z^{H+2}-2(Q-2)\cos\phi\,z^{H+1}+
2(Q-1)z^H+\left(z+Q-2\right)^2.
\]
$P_\phi(z,H,Q)$ is obviously a polynomial for integer values of $H$ and 
(\ref{fzeros}) may be solved numerically. For non-integer values of $H$
the equation (\ref{fzeros}) becomes a transcendental equation and the 
Fisher zeros may be found by numerically checking the condition of existence of 
neutral fixed points of the mapping (\ref{xrec}).

The density of Fisher zeros may be found from (\ref{density}) by substituting
$\xi=z$ and has the form
\begin{equation}
g(z)=\frac1{2\,\pi}\,\frac{HG_1(z,H,Q)-z\,G_2(z,H,Q)}%
{(z-1)\,z\,(z+Q-1)\,\sqrt{-P_0 (z,H,Q)}},
\label{fdensity}
\end{equation}
where
\begin{eqnarray}
G_1(z,H,Q)=(z-1)(z+Q-1)(z^{H+1}-z-Q+2), \nonumber \\
G_2(z,H,Q)=(z^H-1)(\,2\,(z+Q-1)-Q\,z)+Q^2. \nonumber 
\end{eqnarray}
From (\ref{fdensity}) it follows that the density of Fisher zeros is singular
in the Fisher edge singularity points, which are defined as solutions of the equation 
(\ref{fzeros}) for $\phi=0$.

\begin{figure}[h]
\vspace{1cm}
\epsfxsize=14cm
\epsfbox{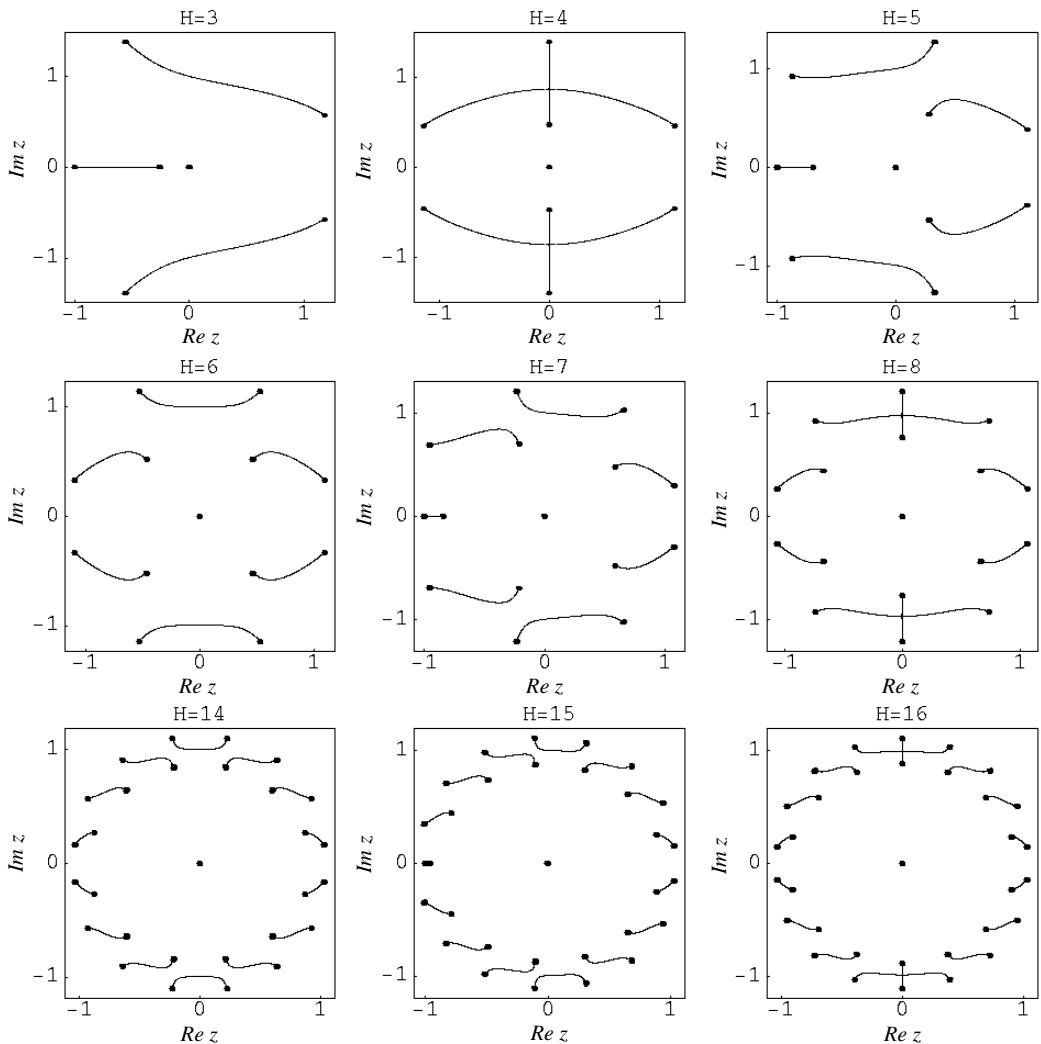}
\caption{The Fisher zeros of 1D Ising model for different values of a magnetic field. 
The big dots at the ends of the lines show Fisher edge singularity points. 
\label{1d:fising1}}
\end{figure}

\begin{figure}[h]
\vspace{1cm}
\epsfxsize=14cm
\epsfbox{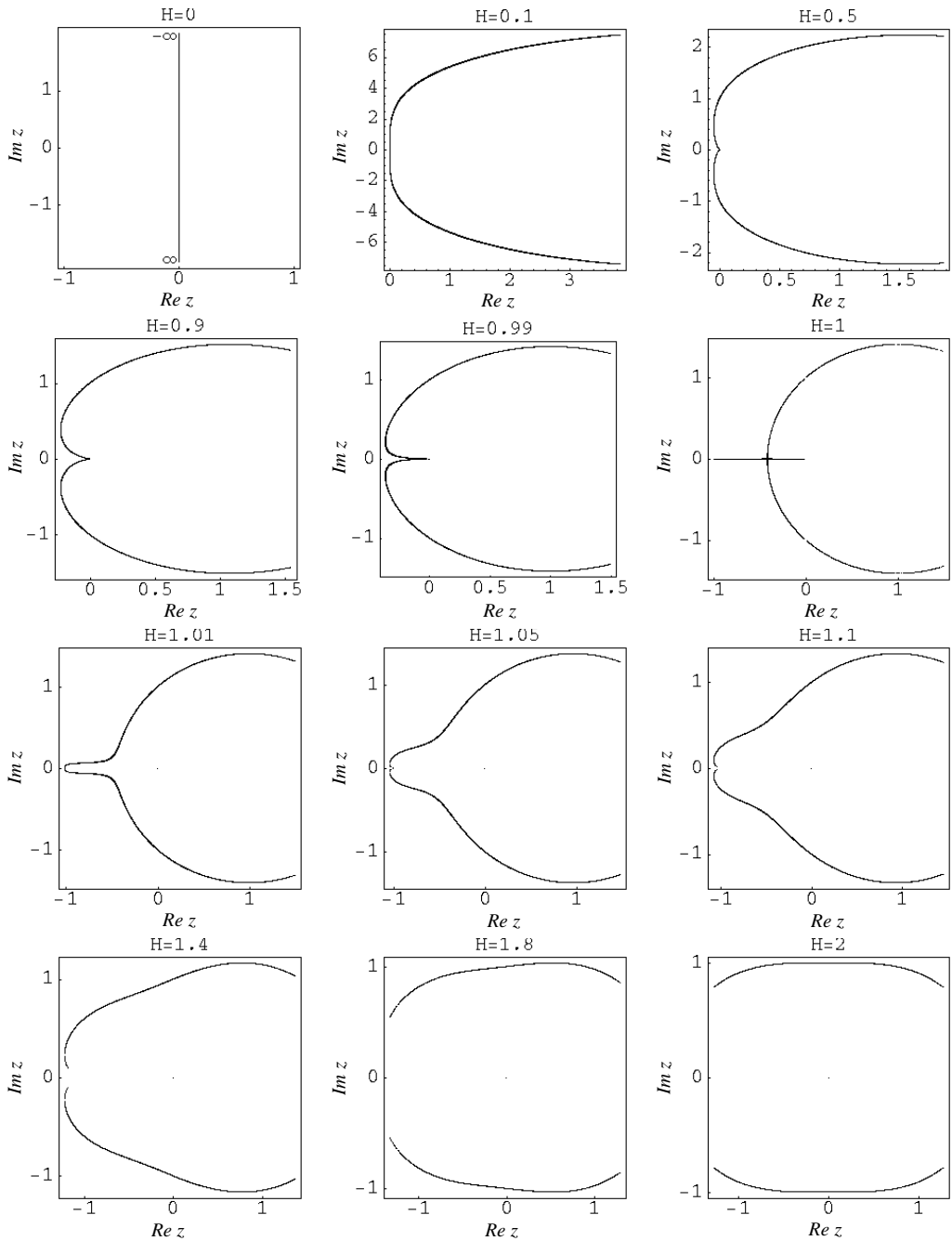}
\caption{The Fisher zeros of 1D Ising model for non-integer values of a magnetic field
$0\leq H\leq 2$.
\label{1d:fising2}}
\end{figure}

Let us first study the Fisher zeros for the Ising model ($Q=2$). For $Q=2$ the 
equation (\ref{fzeros}) has the form
\begin{equation}
z^{2\,H+2}-2\,\cos\phi\,z^{H+2}+4\,\cos^2{\textstyle{\frac\phi 2}}\,z^H+z^2=0.
\label{fisingzeros}
\end{equation}
This equation is symmetric under the $H\rightarrow-H$ transformation 
because of the $Z(2)$ symmetry of the Ising model. Since all coefficients 
of (\ref{fisingzeros}) are real, its solutions will be either real or 
complex conjugate. For integer values of $H$ the Fisher zeros have the form
shown in Figure \ref{1d:fising1}.
From Figure \ref{1d:fising1} and (\ref{fisingzeros}) one can see that the Fisher 
zeros located on lines ending at the Fisher edge singularity points and 
the number of lines equals to the value of the magnetic field $|H|+1$.
Fisher edge singularity points are divided on pairs and every
pair corresponds to a line. Note that $z=0$ is a twice degenerate Fisher edge 
singularity point with the critical exponent $\sigma=-\frac12$. Formally,
two degenerate $z=0$ edge singularity points are considered here as forming a ``line''.
One can see that for odd values of $|H|$ there are negative Fisher zeros and 
for $|H|=4\,n$, $n\in N$, the Fisher zeros are partially located on the imaginary axis. 
Since $P_\phi(z,H,Q)$ is an analytic function of $H$ the location of Fisher zeros 
for non-integer values of $H$ are continuous deformations of Fisher zeros in the 
field $[|H|]$ or $[|H|]+1$, where $[|H|]$ is the integer part of $|H|$, i.e. 
the minimal integer less than $|H|$.
The location of Fisher zeros for non-integer $H$ may be 
found numerically from the condition of existence of neutral fixed points 
of the mapping (\ref{xrec}). As an illustration the Fisher zeros for non-integer 
values of $0<H<2$ are given in Figure \ref{1d:fising2}. 

The Fisher zeros for $Q\neq 2$ Potts model may be studied in the same way 
as for the Ising model. It is impossible to give all possible 
configurations of Fisher zeros for any $H$ and $Q$. Here we give only the 
summary of the main properties. First of all, there is no 
$H\rightarrow -H$ symmetry for $Q\neq 2$ and the Fisher zeros are different 
for positive and negative values of $H$. 

For $Q=1$ the Fisher zeros located on a closed curve and the density of 
Fisher zeros is not singular on this curve. 

For $Q\neq 1$ Fisher zeros are located on several lines and the density 
of Fisher zeros is singular at the end points of these lines with the 
edge singularity exponent $\sigma=-\frac12$.

\begin{figure}[h]
\vspace{1cm}
\epsfxsize=14cm
\epsfbox{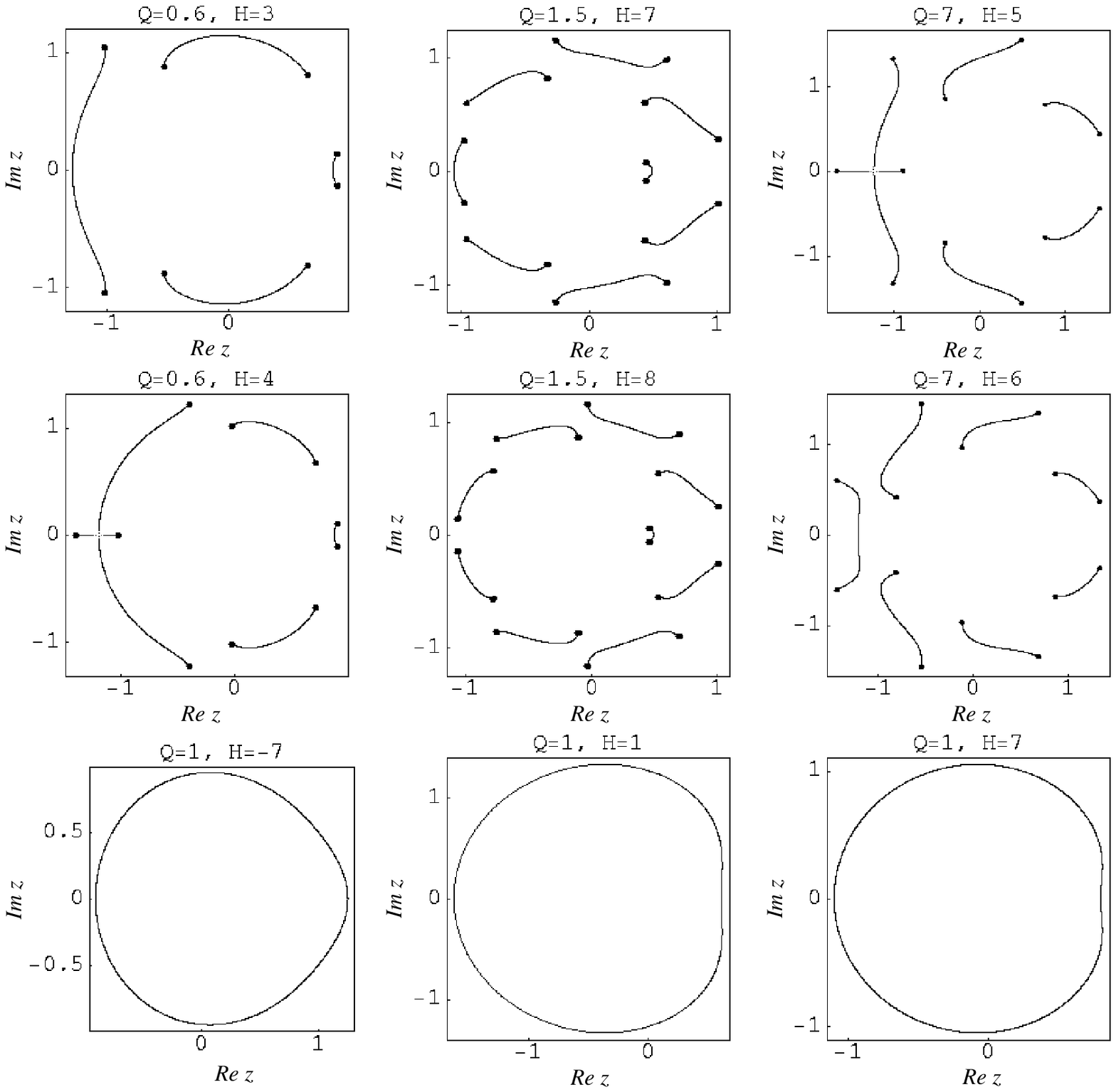}
\caption{The Fisher zeros of the one-dimensional Ising model for some integer 
values of a magnetic field $H$ and a Potts variable $Q$.
\label{1d:fising3}}
\end{figure}

Numerical experiments show that for integer values of $H$ there are 
the following properties: for $0<Q<1$ 
there is an interval of real Fisher zeros only for even values of $H$; for
$1<Q<2$, the Fisher zeros intersect the negative real semi-axis only for odd values
of $H$; for $Q>2$ there is an interval of real Fisher zeros only for odd 
values of $H$. The number of the lines of Fisher zeros are defined as in
the Ising model. In Figure \ref{1d:fising3} we give some plots that illustrate 
these properties.

\section{Conclusions}

In this paper, the Yang-Lee, Fisher and Potts zeros of the one-dimensional 
$Q$-state Potts model are studied using the dynamical systems theory.
A recurrence relation for the partition function is derived. It is shown 
that zeros of the partition function may be associated with neutral fixed 
points of the corresponding recurrence relation. A general equation for zeros 
of the partition function is derived. It is shown that the density of zeros of 
the  partition function is singular in the edge singularity points with
the critical exponent $\sigma=-\frac12$. 

The recursive method used in this paper is equivalent to the transfer matrix 
method, but in some sense it is more flexible. It gives a numerical
algorithm to study the zeros of the partition function. The algorithm is based 
on testing the condition of existence of neutral fixed points of the
corresponding recurrence relation: If neutral fixed points exist, it means that 
the partition function is zero for given values of $z$, $\mu$ and $Q$. In Section 5
 we saw 
that this algorithm is very useful for studying the Fisher zeros in an arbitrary 
magnetic field. Moreover, it is the only method for studying the Fisher zeros 
for non-integer values of $H$, where $H$ is the renormalized magnetic field 
$\mbox{\~{H}}$ and $H=\mbox{\~{H}}/\mbox{\~{J}}$. 

In conclusion we would like to note that the results given in this paper show 
that the thermodynamic properties of the one-dimensional Potts model are completely 
defined by the recurrence relation (\ref{xrec}) or equivalently, by
two non-degenerate eigenvalues of the corresponding transfer matrix (\ref{eigen})
(see also~\cite{Ghulr}). Moreover, it is proved that the phase transition point 
based on dynamical systems approach coincides exactly with the phase transition point 
based on free energy considerations. The dynamical systems approach used here gives a
numerical method which may be used for studying other one-dimensional systems 
for which a one-dimensional recurrence relation may be derived.

\ack
One of the  authors (R G) would like to acknowledge the Abdus Salam Centre for 
Theoretical Physics for hospitality extended during his visit where this work 
was partially performed.

\end{document}